\documentclass[fleqn,10pt,a4paper]{article}
\usepackage{amsmath,amssymb,bm}
\usepackage{cite,epsfig,color,graphicx}

\title{On the entropy of radiation reaction}

\author{D A Burton\thanks{Department of Physics,
Lancaster University, Lancaster, LA1 4YB, UK
and Cockcroft Institute, Daresbury, WA4 4AD, UK.}
\and
A Noble\thanks{Department of Physics,
SUPA and University of Strathclyde, Glasgow, G4 0NG, UK.}
}

\begin{document}
\maketitle
\begin{abstract}
The inexorable development of ever more powerful laser systems has re-ignited interest in electromagnetic radiation reaction and its significance for the collective behaviour of charged matter interacting with intense electromagnetic fields. The classical radiation reaction force on a point electron is non-conservative, and this has led some authors to question the validity of methods used to model ultra-intense laser-matter interactions including radiation reaction. We explain why such concern is unwarranted. 
\end{abstract}

\section{Introduction}
Contemporary advances in ultra-intense laser facilities have driven the recent surge of interest in the collective behaviour of charged matter in extreme conditions, and a particularly fascinating topic in that context concerns the coupling of an electron to its own radiation field~\cite{rohrlich:2007}. An accelerating electron emits electromagnetic radiation, and the energy and momentum carried away by the electromagnetic field must be properly accommodated. In most practical cases, the Lorentz force on an electron due to an applied electromagnetic field is considerably larger than the force due to the electron's emission, and the effect of the recoil due to the emitted radiation is negligible or can be adequately represented using simple physical reasoning. Although such arguments avoid the difficulties that plague more comprehensive analyses, the parameter regimes promised by forthcoming ultra-intense laser facilities ensure that more fundamental considerations are now of practical necessity. For example, ELI~\cite{eli} is expected to operate with intensities $10^{23} {\rm W}/{\rm cm}^2$ and electron energies in the ${\rm GeV}$ range, at which level the radiation reaction force becomes comparable to, and can even exceed, the applied force due to the laser field.

Motivated by experimental developments, recent theoretical work~\cite{tamburini:2011, lehmann:2012} has focussed on the effects of radiation reaction on a bunch of electrons driven by an ultra-intense laser pulse, where the forces between the electrons are negligible compared to the forces exerted by the laser pulse. An outcome of those studies is that the volume of the region of phase space occupied by the bunch reduces with time (the bunch cools) due to radiation reaction. However, the use of kinetic theory to describe a bunch of non-interacting classical point electrons in this context has recently been criticized~\cite{cremaschini:2013} because of the non-Hamiltonian nature of the Landau-Lifshitz equation~\cite{landau:1987} (or its progenitor, the Lorentz-Dirac equation~\cite{dirac:1938}). As a consequence, the entropy $4$-current is not divergenceless in kinetic theories induced from the Landau-Lifshitz equation~\cite{hazeltine:2004, tamburini:2011, lehmann:2012} or from the Lorentz-Dirac equation~\cite{noble:2013}.

Furthermore, inter-particle interactions should not be ignored in all situations where radiation reaction plays a role. Although one might anticipate that the recoil due to emission of radiation will cool the bunch of electrons in all situations, we recently showed~\cite{noble:2013} that inter-particle interactions may heat the bunch. This Letter explores the significance of this observation, and the pathway that we tread leads directly to an explanation of why the recent criticisms given in Ref.~\cite{cremaschini:2013} are unjustified.
\section{Non-relativistic considerations}
The simplest way to quickly obtain a flavour of the effects of inter-particle interactions is to consider the behaviour of a bunch of {\it non-relativistic} electrons, and assume that the inter-particle forces due to the magnetic fields they generate may be neglected. The force on an electron in the bunch is a superposition of the Lorentz forces exerted by the other electrons in the bunch and the force on the electron due to its own radiation field. For simplicity, we neglect collisions between the electrons and represent the inter-particle forces using a mean field approximation $\bm{E}$ to their electric field.

The Abraham-Lorentz equation (see, for example, Ref.~\cite{rohrlich:2007})
\begin{equation}
\label{abraham_lorentz}
m \frac{d^2 \bm{x}}{d t^2} = q\bm{E}(\bm{x},t) + m\tau \frac{d^3 \bm{x}}{d t^3}
\end{equation}
determines the position $\bm{x}(t)$ of a non-relativistic electron in an ambient smooth electric field $\bm{E}$, where $m$ is the mass of the electron, $q=-e$ is the charge on the electron and the time constant $\tau = q^2/6\pi\epsilon_0 m c^3 = 2r_e/3c$ where $r_e$ is the classical radius of the electron.  The total force on the electron is the sum of the mean field approximation $q\bm{E}$ to the total force exerted by other electrons in the bunch and the reaction $m\tau d^3 \bm{x}/d t^3$ due to the electron's own emission. From now on, we will reserve the term {\it bunch} for the smooth continuum specified by the charge density $\epsilon_0 \bm{\nabla}\cdot\bm{E}$.

Following the iterative procedure introduced by Landau and Lifshitz~\cite{landau:1987}, the introduction of the requirement $m\,d^3\bm{x}/dt^3 = q\dot{\bm{E}} + {\cal O}(\tau)$ removes runaway solutions and (\ref{abraham_lorentz}) can be written as
\begin{equation}
\label{non-rel_landau}
m \frac{d^2 \bm{x}}{dt^2} = q\bm{E}(\bm{x},t) + q\tau \bigg[\partial_t \bm{E}(\bm{x},t) + \bigg(\frac{d\bm{x}}{d t} \cdot \bm{\nabla} \bigg) \bm{E}(\bm{x},t)\bigg]
\end{equation}
where $ {\cal O}(\tau^2)$ terms have been dropped and an overdot indicates $d/dt$.

Suppose that the initial position and velocity of the electron are sampled from a statistical ensemble of initial conditions, and let $\langle \bm{x}(t)\rangle$ be the ensemble average of the electron's position at time $t$. Introducing $\bm{x} = \langle\bm{x}\rangle + \bm{\xi}$ into the expansion of (\ref{non-rel_landau}) to leading order in the random variable $\bm{\xi}$ leads to
\begin{equation}
m \frac{d^2 \langle\bm{x}\rangle}{dt^2} = q\bm{E}(\langle\bm{x}\rangle,t) + q\tau \bigg[\partial_t \bm{E}(\langle\bm{x}\rangle,t) + \bigg(\frac{d\langle\bm{x}\rangle}{d t} \cdot \bm{\nabla} \bigg) \bm{E}(\langle\bm{x}\rangle,t)\bigg]
\end{equation}
and
\begin{align}
\label{thermal_kinetic_unsimplified}
\frac{d}{dt}\bigg(\frac{1}{2} m \langle \dot{\bm{\xi}}\cdot{\dot{\bm{\xi}}}\rangle\bigg) =  &\bigg\{q \langle\dot{\xi}^\mu \xi^\nu\rangle \partial_\nu E_\mu
+ q\tau [ \langle\dot{\xi}^\mu \xi^\nu\rangle \partial_\nu \partial_t E_\mu
+ \langle\dot{\xi}^\mu\dot{\xi}^\nu\rangle \partial_\nu E_\mu +\langle\dot{x}^\nu\rangle \partial_\omega\partial_\nu E_\mu \langle\dot{\xi}^\mu \xi^\omega\rangle]\bigg\}\bigg|_{\bm{x}=\langle\bm{x}\rangle}
\end{align}
where Greek indices range over $1,2,3$ and the explicit time dependence of the electric field $\bm{E}$ in (\ref{thermal_kinetic_unsimplified}) has been suppressed for notational convenience.

Simple choices for $\langle\dot{\xi}^\mu \xi^\nu\rangle|_{t=0}$ and $\langle\dot{\xi}^\mu \dot{\xi}^\nu\rangle|_{t=0}$ reveal the significance of (\ref{thermal_kinetic_unsimplified}). Suppose that the initial velocity and initial position of the electron are uncorrelated, and there is no preferred direction for its initial velocity. Hence $\langle \dot{\xi}^\mu \xi^\nu\rangle|_{t=0} = 0$ and $\langle \dot{\xi}^\mu \dot{\xi}^\nu\rangle|_{t=0} = \delta^{\mu\nu} \langle\dot{\bm{\xi}}\cdot\dot{\bm{\xi}}\rangle/3$,  where $\delta^{\mu\nu}$ is the Kronecker delta, and using (\ref{thermal_kinetic_unsimplified}) it follows
\begin{equation}
\label{thermal_kinetic_per_particle_dot}
\frac{d}{dt}\bigg(\frac{1}{2} m \langle \dot{\bm{\xi}}\cdot{\dot{\bm{\xi}}}\rangle\bigg)\bigg|_{t=0} = \bigg[q\tau \frac{1}{3}\langle \dot{\bm{\xi}}\cdot{\dot{\bm{\xi}}}\rangle \bm{\nabla}\cdot\bm{E}\bigg]\bigg|_{\bm{x}=\langle\bm{x}\rangle,\, t=0.}
\end{equation}
 
Let $N$ electrons be represented by a small (finite) element of the bunch, where the element has volume $V$ and the element's centroid is located at $\bm{x}=\langle\bm{x}\rangle$. Hence, the charge density $\rho$  of the bunch and electric field $\bm{E}$ satisfy $\bm{\nabla}\cdot\bm{E} = \rho/\epsilon_0$ with $\rho(\langle\bm{x}\rangle,t) = qN / V$.

If the initial velocities of the $N$ electrons are described by a Maxwell-Boltzmann distribution (with temperature $T$), using (\ref{thermal_kinetic_per_particle_dot}) the thermal kinetic energy $U = N \frac{1}{2} m \langle \dot{\bm{\xi}}\cdot{\dot{\bm{\xi}}}\rangle$ of the $N$ electrons satisfies
\begin{equation}
\label{internal_energy}
\frac{d U}{dt}\bigg|_{t=0} = \bigg[\tau \frac{k_B T}{m\epsilon_0}\rho^2 V\bigg]\bigg|_{\bm{x}=\langle\bm{x}\rangle,\, t=0}
\end{equation}
where $\langle \dot{\bm{\xi}}\cdot{\dot{\bm{\xi}}}\rangle = 3 k_B T/m$ has been used, with $T$ the local temperature of the element. It follows from (\ref{internal_energy}) that $dT/dt|_{t=0} > 0$ and the initial tendency of the element is to heat up, rather than cool down, due to radiation reaction. This result is surprising because we expect the bunch to cool in response to the emission of radiation. 

Although the bunch is not in thermodynamic equilibrium, it is tempting to formally use the first law of thermodynamics $dU = TdS - p dV$ to introduce the entropy $S$ of the element. The volume $V$ of the element satisfies $dV/dt|_{t=0} = 0$ because $V \propto \langle\bm{\xi}\cdot\bm{\xi}\rangle^{3/2}$ and the initial position and velocity of each electron are uncorrelated. Hence, $S$ satisfies
\begin{equation}
\label{non-rel_entropy}
\frac{d S}{dt}\bigg|_{t=0} = \bigg[\tau \frac{k_B}{m\epsilon_0}\rho^2 V\bigg]\bigg|_{\bm{x}=\langle\bm{x}\rangle,\, t=0.}
\end{equation} 
The right-hand side of (\ref{non-rel_entropy}) is strictly positive, which is precisely how one expects the entropy of an isolated bunch of electrons to behave. However, more general considerations show that all is not as it seems.

\section{Relativistic considerations}
The Lorentz-Dirac equation is a fully relativistic description of a structureless point particle in an applied electromagnetic field $F_{ab}$ and has the form
\begin{equation}
\label{LAD}
\frac{d^2 x^a}{d\lambda^2} = -\frac{q}{m} F^a{ }_b\,\frac{dx^b}{d\lambda} + \tau \Delta^a{ }_b \frac{d^3 x^b}{d\lambda^3}
\end{equation}
with $q$ the particle's charge, $m$ the particle's rest mass, $\tau = q^2/6\pi m$ in Heaviside-Lorentz units with $c=\epsilon_0=\mu_0=1$, and the tensor $\Delta^a{ }_b$ is
\begin{equation}
\Delta^a{ }_b = \delta^a_b + \frac{dx^a}{d\lambda} \frac{dx_b}{d\lambda}.
\end{equation}
For an electron, $q=-e<0$ as before. The Einstein summation convention is used throughout the following, indices are raised and lowered using the metric tensor $\eta_{ab} = {\rm diag}(-1,1,1,1)$ and lowercase Latin indices range over $0,1,2,3$. The particle's $4$-velocity $dx^a/d\lambda$ is normalized as follows:
\begin{equation}
\label{proper_time}
\frac{dx^a}{d\lambda} \frac{dx_a}{d\lambda} = -1
\end{equation}
where $\lambda$ is the particle's proper time.

Dirac~\cite{dirac:1938} derived (\ref{LAD}) for a classical point electron by appealing to the conservation condition on the stress-energy-momentum tensor (see Ref. [\citenum{ferris:2011}] for a recent discussion of the derivation). Dirac's approach required a regularization of the electron's singular contribution to the stress-energy-momentum tensor followed by a renormalization of the electron's rest mass. His procedure led to the third-order term in (\ref{LAD}), which is the source of the famous pathological behaviour exhibited by solutions to the Lorentz-Dirac equation (see Ref.~\cite{rohrlich:2007}, and also Ref.~\cite{hammond:2010} for a recent discussion).

The standard approach to ameliorating the problems with the Lorentz-Dirac equation is to replace the third-order terms in (\ref{LAD}) (radiation reaction force) with the derivative of the first term on the right-hand side of (\ref{LAD}) (the applied Lorentz force). This procedure is justifiable if the radiation reaction force is a small perturbation to the Lorentz force, and it yields the Landau-Lifshitz equation~\cite{landau:1987}:
\begin{align}
\label{LL}
\frac{d^2 x^a}{d\lambda^2} = -\frac{q}{m} F^a{ }_b\,\frac{dx^b}{d\lambda} - \tau\frac{q}{m} \partial_c F^a{ }_b \frac{dx^b}{d\lambda} \frac{dx^c}{d\lambda}
+ \tau \frac{q^2}{m^2} \Delta^a{ }_b F^b{ }_c F^c{ }_d \frac{dx^d}{d\lambda}.
\end{align}
Unlike the Lorentz-Dirac equation, the Landau-Lifshitz equation is second order in derivatives in $\lambda$ and its solutions are free from pathologies.

Alternatively, one can derive (\ref{LL}) from a consideration of the stress-energy-momentum balance of an extended charged particle~\cite{gralla:2009}. We will return to this point shortly.

A range of different approaches to modelling the behaviour of a bunch of charged point particles that includes radiation reaction exists in the literature. The most common approach employs the Landau-Lifshitz equation from the outset~\cite{tamburini:2011} , but it is possible to develop a kinetic theory based on the Lorentz-Dirac equation~\cite{noble:2013} that is equivalent to the Landau-Lifshitz kinetic theory to first order in $\tau$. In particular, we showed~\cite{noble:2013} that the entropy $4$-current $s^a$ defined as
\begin{equation}
\label{S_kinetic}
s^a = - k_B \int \dot{x}^a\,g \ln(g)\, \frac{d^3 v}{\sqrt{1+\bm{v}^2}},
\end{equation}
where $k_B$ is Boltzmann's constant, satisfies
\begin{align}
\label{div_S_kinetic}
\partial_a s^a =& -\tau \frac{k_B}{m} \bigg(J_a J^a + 4 \frac{q^2}{m^2} T_{ab} S^{ab}\bigg)
\end{align}
to first order in $\tau$ with
\begin{align}
\label{Ja_g_def}
&J^a = q \int \dot{x}^a g \,\frac{d^3 v}{\sqrt{1+\bm{v}^2}},\\
\label{Sab_g_def}
&S^{ab} = m\int \dot{x}^a \dot{x}^b g\frac{d^3 v}{\sqrt{1+\bm{v}^2}},\\
\label{Tab_def}
&T^{ab} = F^{ac} F^b{ }_c - \frac{1}{4}\eta^{ab} F_{cd} F^{cd}
\end{align}
and $g$ is the $1$-particle distribution of electrons on event-velocity ``phase'' space $(x,\bm{v})$ with  $\dot{x}^\mu = v^\mu$, $\dot{x}^0 = \sqrt{1+\bm{v}^2}$. The vector field $J^a$ is the electric $4$-current of the electron bunch, $S^{ab}$ is the stress-energy-momentum tensor of the electron bunch and $T^{ab}$ is the stress-energy-momentum tensor of the electromagnetic field $F_{ab}$ where
\begin{align}
&\partial_a F^{ab} = J^b,\\
&\partial_a F_{bc} + \partial_b F_{ca} + \partial_c F_{ab} = 0.
\end{align}
Unfortunately, on closer inspection, (\ref{div_S_kinetic}) is an unsettling result. The entropy of any comoving element of an isolated system should not decrease, which in local form is the so-called {\it entropy principle}
\begin{equation}
\label{entropy_principle}
\partial_a s^a \ge 0
\end{equation}
and hence we require
\begin{equation}
\label{J_TS_condition}
J_a J^a + 4\frac{q^2}{m^2} T_{ab} S^{ab} \le 0.
\end{equation}
However, although the right-hand side of the non-relativistic expression (\ref{non-rel_entropy}) is positive, there is no guarantee that (\ref{J_TS_condition}) is satisfied for an isolated bunch. The Maxwell stress-energy-momentum tensor $T^{ab}$ satisfies the energy condition $T_{ab}\, \dot{x}^a \dot{x}^b\ge 0$ at any point $(x,\bm{v})$ and $T_{ab} S^{ab} \ge 0$ immediately follows from (\ref{Sab_g_def}). Although $J^a J_a \le 0$, there is no reason why $J_a J^a$ cannot be overcome by $T_{ab} S^{ab}$ in (\ref{div_S_kinetic}). In general, it seems that (\ref{div_S_kinetic}) cannot describe the evolution of the entropy of an isolated charged bunch.

It is intriguing to note that violations of (\ref{J_TS_condition}) may already be within reach in the laboratory. It has been demonstrated that high-quality femtosecond electron bunches with GeV energies can be created within only a few centimetres or millimetres of laser-plasma, and the opportunities that laser-plasma acceleration offer for the generation of femtosecond X-rays or gamma rays remain a source of intense interest~\cite{schlenvoigt:2008}.  At electron beam energies $\sim 0.1 {\rm GeV}$, the achievable upper limit on the bunch charge is expected to be $\sim 1\,{\rm nC}$ and immediately after exiting the plasma, the bunch in vacuo could have a width $\sim 1 \,{\rm \mu m}$ and length $\sim 1\,{\rm \mu m}$ in the laboratory frame~\cite{gruner:2007}. The electrostatic repulsion within the bunch is very strong and space-charge effects are considerable~\cite{gruner:2007}. 

Due to relativistic effects, the length of the bunch in its instantaneous rest frame will be much greater than its width. Let $L$ be the length of a homogenous cylindrical bunch of electrons, let $R$ be its radius, with $R \ll L$, and let $V^a = J^a/q n$ be the $4$-velocity of the bunch with $q n = -\sqrt{-J^a J_a}$ the proper charge density of the bunch. Neglecting effects due to the finite length of the bunch, the electric field inside the bunch satisfies $|\bm{E}| \approx -q n r/2$ at radial distance $r$ from the bunch's axis of symmetry. Neglecting the thermal spread of the $1$-particle distribution $g$ gives $S^{ab} \approx mn V^a V^b$, and neglecting the magnetic fields generated by the bunch yields $T_{ab} S^{ab} \approx m n {\cal E}$ where ${\cal E} = T_{ab} V^a V^b \approx \bm{E}^2/2$ is the energy density of the electromagnetic field in the instantaneous rest frame of the bunch. Thus, (\ref{J_TS_condition}) leads to
\begin{equation}
\label{energy_density_condition}
{\cal E} \lesssim m n/4
\end{equation}
which evaluated at $r=R$ gives
\begin{equation}
\label{L_bound}
\frac{Q^2}{2\pi M} \lesssim L
\end{equation}
where $Q$ is the charge of the bunch and $M$ is its mass. It follows from (\ref{L_bound}) that the number $N$ of electrons comprising the bunch is bounded from above:
\begin{equation}
\label{N_bound}
N \lesssim \frac{L}{2 r_e}
\end{equation}
where $r_e$ is the classical radius of the electron.

Using the value $L= 0.26\, {\rm mm}$ given in Ref.~\cite{gruner:2007} for the length of the bunch in its instantaneous rest frame, equation (\ref{N_bound}) yields
\begin{equation}
N \lesssim 4.6 \times 10^{10}
\end{equation} 
which corresponds to the bound $Q \lesssim 7.4\, {\rm nC}$ and is within an order of magnitude of the achievable values specified in Ref.\cite{gruner:2007}. Hence, it is possible that the bunch will violate (\ref{energy_density_condition}) outside the plasma in regions where externally applied fields are negligible. Of course, any violation of (\ref{entropy_principle}) can only last for a very short time; the bunch will undergo a ``transverse Coulomb explosion''~\cite{gruner:2007} and its radius will increase by about $2$ orders of magnitude over a time interval of $1\, {\rm ps}$.

A particularly intriguing conclusion is obtained when (\ref{J_TS_condition}) is applied to a spherically symmetric and homogeneous bunch of {\it cold} electrons. In this case, the spherical symmetry ensures that the magnetic field vanishes and the electromagnetic energy density ${\cal E} = T_{ab} V^a V^a$ of the bunch satisfies ${\cal E} = \bm{E}^2/2$ exactly. The electric field $\bm{E}$ is purely radial and has magnitude $|\bm{E}| = - q n r/3$ inside the bunch, where $r$ is the distance from the centre of the bunch in its rest frame. Equation (\ref{J_TS_condition}) leads to ${\cal E} \leq m n/4$, which evaluated at $r=R$ yields
\begin{equation}
\label{QMR_bound}
\frac{Q^2}{6\pi M} \le R
\end{equation}
where $Q$ is the charge of the bunch, $M$ is its mass and $R$ is its radius. Recalling that we have used units in which $c=\epsilon_0=\mu_0=1$, it is interesting to note that the factor $4$ in (\ref{J_TS_condition}) ensures that the bound (\ref{QMR_bound}) is saturated by an expression identical to $\tau = q^2/6\pi\epsilon_0 mc^3$ under the replacement $(q,m) \mapsto (Q,M)$.

\section{Stress-energy-momentum conservation}
The failure of $s^a$ to satisfy the entropy principle may be resolved by appealing to the dynamics of a system of classical {\it extended} charged particles. The equation of motion of an extended particle must be compatible with stress-energy-momentum conservation
\begin{equation}
\partial_a (s^{ab} + t^{ab}) = 0 
\end{equation}
where $t^{ab}$ is the stress-energy-momentum of the electromagnetic field $f_{ab}$,
\begin{equation}
\label{particle_EM_stress}
t^{ab} = f^{ac} f^b{ }_c - \frac{1}{4}\eta^{ab} f_{cd} f^{cd},
\end{equation}
and $s^{ab}$ the stress-energy-momentum tensor of the particle.
The electromagnetic field $f_{ab}$ satisfies Maxwell's equations
\begin{align}
& \partial_a f^{ab} = j^b,\\
& \partial_a f_{bc} + \partial_b f_{ca} + \partial_c f_{ab} = 0
\end{align}
with $j^b$ the electric $4$-current of the particle. The electromagnetic field is decomposed as $f^{ab} = f^{ab}_{\rm ext} + f^{ab}_{\rm self}$ where the external field $f^{ab}_{\rm ext}$ is generated by sources other than the particle and satisfies the vacuum Maxwell equation $\partial_a f^{ab}_{\rm ext} = 0$, and the particle's self-field $f^{ab}_{\rm self}$ satisfies $\partial_a f^{ab}_{\rm self} = j^b$. The stress-energy-momentum tensor $t_{ab}$ is quadratic in $f_{ab}$ and may be decomposed as $t^{ab} = t^{ab}_{\rm ext} + t^{ab}_{\rm self} + t^{ab}_{\rm cross}$ where $t^{ab}_{\rm ext}$ (resp. $t^{ab}_{\rm self}$) is (\ref{particle_EM_stress}) with $f^{ab}$ replaced by $f^{ab}_{\rm ext}$ (resp. $f^{ab}_{\rm self}$). The remaining term $t^{ab}_{\rm cross}$ arises because (\ref{particle_EM_stress}) is quadratic in $f_{ab}$. It may be shown that $\partial_a (t^{ab}_{\rm ext} + t^{ab}_{\rm cross}) = f^{bc}_{\rm ext} j_c$ and hence
\begin{equation}
\label{stress_balance_extended_charge}
\partial_a(s^{ab} + t^{ab}_{\rm self}) = - f^{bc}_{\rm ext} j_c.
\end{equation}
The Landau-Lifshitz equation for a point particle may be obtained from (\ref{stress_balance_extended_charge}) by requiring that the fields of the extended particle behave in a prescribed manner under a particular one-parameter family of transformations that shrinks the world tube of the extended particle down to the world line of the point particle~\cite{gralla:2009}. This process requires a renormalization of the mass of the point particle corresponding to a re-identification $s^{ab} \rightarrow s^{\prime ab}$ of the extended particle's stress-energy-momentum tensor. Hence
\begin{equation}
\label{renormalized_stress_balance_extended_charge}
\partial_a s^{\prime ab} = - f^{bc}_{\rm ext} j_c - \partial_a t_{\rm self}^{\prime ab}
\end{equation}
where $t_{\rm self}^{\prime ab} = t^{ab}_{\rm self} + s^{ab} - s^{\prime ab}$ and, unlike (\ref{particle_EM_stress}), $t_{\rm self}^{\prime ab}$ is generally not traceless. 

It is straightforward to generalize (\ref{renormalized_stress_balance_extended_charge}) to describe the stress-energy-momentum balance of a collection of extended charged particles with non-intersecting world tubes. It follows
\begin{equation}
\label{renormalized_stress_balance_extended_multi_charge}
\partial_a \bigg(\sum\limits_N s_N^{\prime ab}\bigg) = - \sum\limits_N f^{bc}_{N\,{\rm ext}} j_{N c} - \partial_a \bigg(\sum\limits_N t_{N\, {\rm self}}^{\prime ab}\bigg)
\end{equation}
where $f^{ab}_{N\, {\rm ext}} = \sum_{M \neq N} f^{ab}_{M\, {\rm self}}$ satisfies
\begin{equation}
\partial_a f^{ab}_{N\, {\rm ext}} = \sum\limits_{M \neq N} j^b_M.
\end{equation}
Each value of the index $N$ corresponds to a different extended particle and the supports of the $4$-currents $j^a_N$, $j^a_M$ (with $N\neq M$) do not intersect.

The initial supports of the particles' world tubes are specified as the intersections of the world tubes with a fiducial spacelike hypersurface, and a system of field equations for a bunch of charged extended particles is obtained using an ensemble average $\langle\dots\rangle$ over the initial supports. The details of the probability distribution are not required for present purposes.

The total $4$-current may be expressed as $\sum_N j^a_N = J^a + \delta j^a$ where $J^a = \sum_N \langle j^a_N \rangle$ and the fluctuation $\delta j^a$ satisfies $\langle \delta j^a\rangle = 0$. Hence
\begin{align*}
&\partial_a \langle f^{ab}_{N\, {\rm ext}}\rangle = J^a - \langle j^a_N \rangle,\\
&\partial_a \langle f_{bc}^{N\, {\rm ext}}\rangle + \partial_b \langle f_{ca}^{N\, {\rm ext}}\rangle + \partial_c \langle f_{ab}^{N\, {\rm ext}}\rangle = 0.
\end{align*}
Thus, the electromagnetic field $f^{ab}_{N\,{\rm ext}}$ external to the $N{\rm th}$ extended particle may be decomposed as $f^{ab}_{N\,{\rm ext}} = F^{ab} + \delta f^{ab}_{N\,{\rm ext}}$ where
\begin{align}
&\partial_a \langle \delta f^{ab}_{N\,{\rm ext}}\rangle = - \langle j^b_N\rangle,\\
&\partial_a \langle \delta f_{bc}^{N\,{\rm ext}}\rangle + \partial_b \langle \delta f_{ca}^{N\,{\rm ext}}\rangle + \partial_c \langle \delta f_{ab}^{N\,{\rm ext}}\rangle = 0
\end{align}
and 
\begin{align}
\label{maxwell_smeared}
&\partial_a F^{ab} = J^b,\\
\label{bianchi_smeared}
&\partial_a F_{bc} + \partial_b F_{ca} +  \partial_c F_{ab} = 0.
\end{align}
Using (\ref{renormalized_stress_balance_extended_multi_charge}) it follows
\begin{equation}
\partial_a S^{ab} = - F^b{ }_c\, J^c - \partial_a \Pi^{ab} - \sum\limits_N \langle \delta f^{bc}_{N{\rm ext}} j_{N c}\rangle 
\end{equation}
where $S^{ab} = \sum\limits_N \langle s_N^{\prime ab}\rangle$ is identified as the stress-energy-momentum tensor of the bunch and $\Pi^{ab} = \sum\limits_N \langle t_{N\, {\rm self}}^{\prime ab}\rangle$ is a remnant of the total self-field stress-energy-momentum of the extended particles. Assuming that the correlation between the fluctuation $\delta f^{ab}_{N{\rm ext}}$  of the field external to each extended particle and that particle's $4$-current $j^a_N$ are negligible relative to the coarse-grained self-force $\partial_a \Pi^{ab}$ of the bunch, we obtain
\begin{align}
\label{stress_balance_smeared}
\partial_a S^{ab} = - F^{bc} J_c - \partial_a \Pi^{ab}.
\end{align}
It is clear from the above that the entropy of the bunch must include a contribution arising from the remnant $\Pi^{ab}$ due to the self-fields of the extended particles, and that contribution is missing from (\ref{S_kinetic}). We see that the entropy of the bunch should be redefined as
\begin{equation}
\label{redefined_sa}
s^a = - k_B \int \dot{x}^a\,g \ln(g)\, \frac{d^3 v}{\sqrt{1+\bm{v}^2}} + \sigma^a
\end{equation}
where the divergence of the entropy $4$-current $\sigma^a$ associated with $\Pi^{ab}$ compensates the divergence of the first term in (\ref{redefined_sa}) and yields $\partial_a s^a \ge 0$ overall for an isolated system.

The Landau-Lifshitz equation may be recovered from (\ref{stress_balance_extended_charge}) in the limit as the extended particle is shrunk to a point charge~\cite{gralla:2009}, and this result motivates our assertion that (\ref{maxwell_smeared}, \ref{bianchi_smeared}, \ref{stress_balance_smeared}) is a valid description of a bunch of point electrons with $J^a$, $S^{ab}$ specified by (\ref{Ja_g_def}, \ref{Sab_g_def}) and with $\Pi^{ab}$ chosen appropriately. In this case, the ``hidden'' entropy current $\sigma^a$ is expected to capture a flavour of the disorder in the near-zone fields of the electrons. Although a full analysis of the properties of $\sigma^a$ and $\Pi^{ab}$ is beyond the scope of the present article, it is already clear that there is no need to jettison the kinetic theories derived from the Landau-Lifshitz or Lorentz-Dirac equations as suggested recently in Ref.~\cite{cremaschini:2013}. The above shows that the Vlasov equation presented in Ref.~\cite{noble:2013} is no less consistent than the usual Vlasov equation derived from the Lorentz force in which radiation reaction is neglected.
\section*{Acknowledgements}
DAB would like to thank Robin W Tucker for inspiring this article. This work was undertaken as part of the ALPHA-X consortium funded under EPSRC grant EP/J018171/1. DAB is also supported by the Cockcroft Institute of Accelerator Science and Technology  (STFC grant ST/G008248/1).
%


\begin{thebibliography}{9}
\bibitem{rohrlich:2007} F. Rohrlich, Classical Charged Particles, second ed., World Scientific, 2007.
\bibitem{eli} http://www.extreme-light-infrastructure.eu/
\bibitem{tamburini:2011} M. Tamburini {\it et al.}, Nucl. Instrum. Methods A 653 (1) (2011) 181. 
\bibitem{lehmann:2012} G. Lehmann and K.H. Spatschek, Phys. Rev. E 85 (2012) 056412.
\bibitem{cremaschini:2013} C. Cremaschini and M. Tessarotto, Phys. Rev. E 87 (2013) 032107.
\bibitem{landau:1987} L.D. Landau and E.M. Lifshitz, The Classical Theory of Fields (Course of Theoretical Physics, vol. 2), fourth ed., Butterworth-Heinemann Ltd, 1987.
\bibitem{dirac:1938} P.A.M. Dirac, Proc. Roy. Soc. A 167 (1938) 148.
\bibitem{hazeltine:2004} R.D. Hazeltine and S.M. Mahajan, Phys. Rev. E 70 (2004) 046407.
\bibitem{noble:2013} A. Noble, D.A. Burton, J. Gratus and D.A. Jaroszynski,  J. Math. Phys. 54 (2013) 043101.
\bibitem{ferris:2011} M.R. Ferris and J. Gratus, J. Math. Phys. 52 (2011) 092902.
\bibitem{hammond:2010} R.T. Hammond, Phys. Rev. A 81 (2010) 062104.
\bibitem{gralla:2009} S.E. Gralla, A.I. Harte and R.M. Wald, Phys. Rev. D 80 (2009) 024031.
\bibitem{schlenvoigt:2008} H.P. Schlenvoigt {\it et al}, Nat. Phys. 4 (2008) 133.
\bibitem{gruner:2007} F. Gr\"uner, {\it et al}, Appl. Phys. B 86 (2007) 431.
\end{thebibliography}
\end{document}